\documentclass[12pt]{article}
\usepackage{amsmath,amssymb,amsthm,color}
\usepackage{graphicx}
\usepackage{cite}
\usepackage{epsfig}
\usepackage{multirow}
\renewcommand{\baselinestretch}{2}
\begin{document}
%
\title{Configuration- and concentration-dependent electronic properties of hydrogenated graphene\\}
\author{
\small Hao-Chun Huang$^{a}$, Shih-Yang Lin$^{a,*}$, Chung-Lin Wu,$^{a}$ Ming-Fa Lin$^{a,*}$ $$\\
\small $^a$Department of Physics, National Cheng Kung University, Tainan 701, Taiwan \\
 }
\renewcommand{\baselinestretch}{1}
\maketitle

\renewcommand{\baselinestretch}{1.4}
\begin{abstract}

The electronic properties of hydrogenated graphenes are investigated with the first-principles calculations. Geometric structures, energy bands, charge distributions, and density of states (DOS) strongly depend on the different configurations and concentrations of hydrogen adatoms. Among three types of optimized periodical configurations, only in the zigzag systems the band gaps can be remarkably modulated by H-concentrations. There exist middle-gap semiconductors, narrow-gap semiconductors, and gapless systems. The band structures exhibit the rich features, including the destruction or recovery of the Dirac-cone structure, newly formed critical points, weakly dispersive bands, and (C,H)-related partially flat bands. The orbital-projected DOS are evidenced by the low-energy prominent peaks, delta-function-like peaks, discontinuous shoulders, and logarithmically divergent peaks. The DOS and spatial charge distributions clearly indicate that the critical bondings in C-C and C-H is responsible for the diversified properties.
\vskip 1.0 truecm
\par\noindent

\vskip1.0 truecm

\par\noindent  * Corresponding authors. {~ Tel:~ +886-6-2757575-65272 (M.F. Lin)}\\~{{\it E-mail address}: mflin@mail.ncku.edu.tw (M.F. Lin), clwuphys@mail.ncku.edu.tw (C.L. Wu)}
\end{abstract}
\pagebreak
\renewcommand{\baselinestretch}{2}
\newpage

{\bf 1. Introduction}
\vskip 0.3 truecm

Graphene, a single layer of graphite, is an allotrope of two-dimensional carbon structures with hexagonal lattices \cite{novoselove2005two}. It has attracted a lot of considerable theoretical and experimental researches \cite{schedin2007detection,geim2009graphene}, mainly related to the remarkably essential properties, such as the ultrahigh carrier mobility \cite{berger2006electronic}, superior thermal conductivity \cite{balandin2008superior}, and quite large tensile strength \cite{lee2008measurement}, and so on. Graphene has an isotropic Dirac-cone structure, while it is a zero-gap semiconductor because of the vanishing density of states at the Fermi level (EF=0). The electronic properties can be significantly modulated by the adatom dopings \cite{liu2011chemical,wei2009synthesis,nakada2011migration}, graphene layers \cite{sutter2009electronic,lauffer2008atomic}, stacking configurations \cite{zhong2012stacking,tran2015configuration}, external fields \cite{lai2008magnetoelectronic,lin2015magneto}, and mechanical strains \cite{wong2012strain,pereira2009strain}. Among the above-mentioned methods, atom adsorption is the most effective way to dramatically induce the metallic or semi-conducting behavior. For example, lithiated graphene is a metal while oxygenated graphene is a semiconductor. In this work, the first-principles calculations are used to comprehend how to modulate the electronic properties of hydrogenated graphene by the different distributions and concentrations. The close relations between energy bands and orbital hybridizations in chemical bonds (orbital bondings) will be investigated in detail.

Hydrogenated graphene, which first reported by Elias et al. \cite{elias2009control}, showed that the electronic properties could be controlled by H adsorption and desorption. This attracts a bunch of experiments studies, including uniform or non-uniform distribution of H atoms. Balog et al. \cite{balog2010bandgap} observed that patterned H adsorption on the Moir\'{e} superlattice positions of graphene can open a band gap of 0.45 eV. The STM image presented by Sessi et al. \cite{sessi2009pattering} displays the disordered morphology of a hydrogen-saturated surface. As for the H-concentration, it can be controlled by the exposure time to hydrogen plasma, and verified by the intensity ratio of D and G band in Raman spectroscopy \cite{Jaiswal2011controlled}. Most importantly, the hydrogen atom-induced gap is successfully observed by angle-resolved photoemission spectroscopy (ARPES) \cite{balog2010bandgap} and scanning tunneling spectroscopy (STS) experiments \cite{castellanos2012reversible}, being useful for its sufficient size to electronic applications at room temperature.

Unlike pristine graphene, the hydrogenated grphene is not stable as a perfectly planar sheet. Because the sp$^3$ hybridization is more stable than the sp$^2$ one, it is energetically favorable as a low buckled structure. Recently, their electronic structure of fully hydrogenated ones has been revealed to be insulators by the density
functional theory (DFT) calculations \cite{sofo2007graphane}. Other theoretical works exploring the disappearance of band gap under specific hydrogen distribution \cite{chandrachud2010a}. Various H-concentration and distribution are also studied with diversified H arrangements \cite{balog2010bandgap,yang2010two,gao2011band}. These results suggest that the electronic properties can be altered drastically. However, The important orbital hybridizations in C-H and C-C bonds are not fully analyzed, and the three prototype of hydrogen configurations with various concentration has not been systematically investigated.

In this paper, the geometric and electronic properties of hydrogenated graphene are investigated by using the first-principles calculations. The relations with various arrangements and concentrations of hydrogen atoms are taken into account. The calculations on bond lengths, bond angles, atom heights, energy dispersions, band-edge states, band gaps, charge density, and density of states (DOS) are explored in detail. The spatial charge distribution and the orbital-projected DOS are useful to comprehend how the electronic properties are strongly affected by the chemical bondings between two atomic orbitals. Apparently, this work shows the diversified electronic band structures, including the presence or the absence of the Dirac-cone structure, the anisotropic energy bands, the anti-symmetric valence and conduction bands, three kinds of energy dispersions near the Fermi level (E${_F}$=0), the gapped or gapless behavior, and a lot of band-edge states. Such predictions are further reflected in the feature-rich DOS. The predicted atomic positions, energy bands, and DOS could be verified by scanning tunneling microscopy (STM), APRES, and STS, respectively.\\

\vskip 0.6 truecm
\par\noindent
{\bf 2. Methods }
\vskip 0.3 truecm

The first-principles calculations on the geometric and electronic properties of hydrogenated graphenes are performed within the density functional theory implemented by using the Vienna \emph{ab initio} simulation package (VASP) \cite{kresse1996efficient,kresse1999ultrasoft}. The electron-electron interactions are evaluated from the exchange-correlation function under the generalized gradient approximation of Perdew-Burke-Ernzerhof \cite{perdew1996generalized}, whereas the electron-ion interactions are calculated using the projector augmented wave method \cite{blochl1994projector}. The wave functions are built from a plane-wave set with a maximum energy cutoff of 500 eV. To avoid interactions between adjacent unit cells, a vacuum space of 10 $\AA$ is inserted between replicate systems. For band-structure calculations and geometry relaxation and band-structure calculations, the first Brillouin zones is sampled by $78\times78\times1$ and $14\times14\times1$ \emph{k}-points in the Monkhorst-Pack scheme, respectively. The convergence criterion of the Helmann-Feymann force is set to be less than 0.01 eV/\AA.\\

\vskip 0.6 truecm
\par\noindent
{\bf 3. Results and discussion}
\vskip 0.3 truecm

We utilize the density-functional theory to systematically investigate the optimized geometric structures of hydrogenated graphene, including three classical configurations with different hydrogen distributions, as what had been done for the geometric characterization of carbon nanotube (Fig. 1(a)). In the case of fully hydrogenated graphene (graphane), each unit cell contains two carbons and two hydrogens, with the latter being located at the top and the bottom of the carbon plane, as shown in Fig. 1(b) for the chair-like structure. As the hydrogen concentration decreases, the geometric structures can be classified into three types, based on the arrangement of carbon atoms between two hydrogen atoms. The first type is the zigzag-hydrogenated graphene, containing $100\%$, $25\%$ (Fig. 1(c)), $11.1\%$, $6.3\%$, $4\%$, and $2\%$ H-concentrations, respectively, with the hydrogen-carbon ratios, $2:2$, $2:8$, $2:18$, $2:32$, $2:50$, and $2:98$. It is noticed that a primitive unit cell is enlarged as the concentration decreases. The second type, namely the armchair-hydrogenated graphene, possesses its concentrations diminished as $33\%$ (Fig. 1(d)), $8.3\%$, and $3.7\%$ where the hydrogen-carbon ratios are $2:6$, $2:24$, and $2:54$, respectively. As for the third type of chiral-hydrogenated graphene, the H-C ratios are $2:14$, $2:56$ and $2:126$, corresponding to $14.3\%$ (Fig. 1(e)), $3.6\%$ and $1.6\%$ H-concentrations, respectively. All the three types reveal similar geometric deformations; however, these three prototypes may lead to various feature-rich electronic structures.

The main characteristics of geometric structures, including the C-C bond lengths, H-C bond lengths, H-C-C angles, and heights of carbon atoms, strongly depend on the concentration and distribution of hydrogen atoms, as shown in Table 1. All the C-C bond lengths in graphene are 1.41 {\AA}, while they become non-uniform in hydrogenated graphene. The nearest and the next-nearest C-C bond lengths, measured from the H atom, are in the ranges of 1.47 $\rightarrow$ 1.53 {\AA} and 1.36 $\rightarrow$ 1.40 {\AA}, respectively. As to the H-C bond lengths, they are hardly influenced by hydrogen concentration, in which they keep in the range of 1.12 $\rightarrow$ 1.15 {\AA}. The H-C-C angles present a significant change from 100.5$^{\circ}$ to 107.1$^{\circ}$. With the diluted H-concentration, the H-C-C angles in chiral configuration decrease, but the opposite is true for those in armchair one. Furthermore, a simple relation is absent for zigzag configuration. On the other hand, the concentration-dependence of the C-C-C angles on is different from that of the H-C-C angles. Apparently, the carbon atoms nearest to the hydrogen atoms deviate from the plane, in which their heights grow from 0.22 to 0.57 {\AA} in the decrease of H-concentration. The shift of height is greater as the concentration dips.

The above-mentioned geometric characteristics are closely related to the strong C-C and C-H chemical bondings. The latter are formed by the $sp^3$ hybridization of four orbitals (2$s$, 2$p_x$, 2$p_y$, 2$p_z$) and the hybridization of $sp^3$ and 1$s$ orbitals (Fig. 2(a)). The planar C-C bond lengths are mainly determined by the $\sigma$ bonding due to the (2$s$, 2$p_x$, 2$p_y$) orbitals (Fig. 2(b)). These orbitals are partially participated in the bonding between C and H atoms. Consequently, the weakened $\sigma$ bonding on the plane has an effect to lengthen the adjacent C-C bond lengths. For the C-C bonds next to these lengthened C-C bonds, their lengths are actually shortened. Moreover, the bidirectional C-H bondings above and below the graphene plane can induce the planar strain. With the decreasing H-concentration, the strain forces are balanced by more carbon atoms in a unit cell, being the main reason for the larger displacements of the nearest-neighbor carbons.

The experimental measurements by STM can provide the spatially atomic distributions at the local nano-structures. They have been successfully utilized to resolve the unique geometric structures of the graphene-related systems, such as, graphene \cite{de2008periodically}, graphene compounds \cite{balog2007bandgap}, graphite \cite{cervenka2009room}, graphene nanoribbons \cite{ruffieux2012electronic} and carbon nanotubes \cite{wilder1998electronic}. The atomic-scaled observations clearly show the rippled and buckled structures \cite{de2008periodically}, the adatom distributions on graphene surface \cite{kondo2012atomic}, the local defects \cite{cervenka2009room}, and the chiral arrangements of the hexagons on the planar edges \cite{tao2011spatially} and a cylindrical surface \cite{wilder1998electronic}. The predicted geometric properties in hydrogenated graphene, which include the almost unchanged H-C bond lengths, and the significant changes in the C-C bond lengths, the carbon height and the H-C-C angles, deserve a closer experimental examination. Also, this measurement is useful in understanding the chemical bondings between H and C atoms.

The electronic structure of hydrogenated graphene is highly enriched by the strong orbital hybridization in C-H bonds. For graphene, the $sp^{2}$ orbitals of ($2p_{x}$, $2p_{y}$, 2$s$) possess the very strong covalent $\sigma$-bondings among three nearest-neighboring carbon atoms, and the 2$p_z$ orbitals contribute to the $\pi$-bondings (Fig. 2(c)), being associated with the highest occupied valence band and the lowest unoccupied conduction band. The low-lying valence and conduction bands, characterized by the isotropic linear bands intersecting at the K point, purely come from the $\pi$-bonding of 2$p_z$ orbitals. As the state energy increases, they gradually become parabolic energy dispersions, in which there are special saddle points at the M point with energies $-$2.2 and 1.8 eV.
The two middle-energy $\sigma$ bands of $2p_{x}$ and $2p_{y}$ orbitals are formed with the initial energy $-3$ eV at the $\Gamma$ point and extended with the variation of wave vector. Moreover, the deepest $\sigma$ band, with energy $-19.5$ eV at the $\Gamma$ point, is dominated by the 2$s$ orbitals. Those three energy levels of $\sigma$ and $\pi$ bands constitute the whole band structure of graphene.

The band structure is dramatically changed by the strong H-C chemical bondings. All the energy bands are anisotropic along the high-symmetry points; furthermore, valence bands are highly asymmetric to the conduction bands about the Fermi level ($E_{F}=0$), as shown in Figs. 3 and 4. There exist three kinds of energy dispersions, including linear, parabolic and partial flat ones. The linear isotropic Dirac cone of graphene, as shown in Fig. 3(a), is destroyed in most concentrations, mainly owing to the absence of the hexagonal symmetry. The band-edge states, which correspond to the extreme or the saddle points in the energy-wave-vector space are, in general, presented at the highly symmetric points (K, $\Gamma$ and M), but are also revealed at the other wave vectors for a specific concentration. Furthermore, the hydrogenated system might have a middle, narrow or vanishing energy gap, depending on the distribution and concentration of hydrogen.

Energy gaps of zigzag-hydrogenated graphenes could be tuned by the H-concentration. The fully hydrogenated system exhibits the largest direct energy gap of E$_g$ = 3.5 eV at the $\Gamma$ point, (Fig. 3(b)), in which the highest occupied state (HOS; E$^{v}$ = $-$1.7 eV) and the lowest unoccupied state (LUS; E$^{c}$ = 1.7 eV), respectively, correspond to the $\sigma$ bands of $2p_{x}$ and $2p_{y}$ orbitals and the $\pi$* band of $2p_{z}$ orbitals. The $\pi$ band shifts to energy deeper than -3.5 eV; furthermore, hydrogen atoms make main contribution to the energy bands in the range of E$^{v}$ $\sim$ $-$3.5 $\le$ E $\le$ $-$8 eV (blue circles). The strong H-C bondings and the very low bound state energy of H atom are responsible for the drastic changes in low-lying bands. Those features are similar to that of fully hydrogenated silicene. As the concentration decreases, the $\pi$ bands gradually approach to the Fermi level, but the opposite is true for the $\sigma$ bands. For the 25$\%$ system, the highest occupied state coming from the $\pi$ band shifts to the M point and thus creates an in-direct gap of E$_g$ = 2.5 eV. The 6.3$\%$ concentration shows a 0.25 eV direct band gap at the $\Gamma$ point and a partially flat bands at around $-$2.5 eV. Peculiarly, for the 4$\%$ concentration, the highest occupied valence state and the lowest unoccupied conduction state occur at M and K points respectively, leading to an indirect band gap of 0.75 eV. The 2$\%$ concentration possesses the smallest direct band gap of 0.2 eV at the $\Gamma$ point, being determined by non-passivated C atoms. In general, direct or indirect H-dependent energy gaps decline fluctuated, and coming from the gradually recovery of $\pi$ bonding (Fig. 5).

Electronic properties are sensitive to the changes in the distribution of hydrogen. For the armchair hydrogenated systems, there exist two weakly dispersive energy bands across E$_F$, as indicated in Figs. 4(a)-4(c). They come from the partially localized electrons of H atoms and the second-, fourth-, or sixth-nearest C atoms (not shown). All of them exhibit the gapless behavior for any H-concentrations. The energy dispersions become weaker in the descending order of the 33$\%$, 8.3$\%$, and 3.7$\%$ H-concentrations, and the corresponding band widths are 1.2 eV, 0.3 eV, and 0.1 eV.
On the other hand, the energy bands dominated by the other non-passivated C atoms reveal the local minima (maxima) at $+$2 eV and $+$0.5 eV ($-$1.8 eV and $-$0.9 eV) at the $\Gamma$ point for the 33$\%$ and 8.3$\%$ H-concentration, respectively. Specially, the distorted Dirac cone is recovered at the $\Gamma$ point at the sufficiently low concentration (3.7$\%$), as shown in Fig. 4(h). Furthermore, the H atoms and passivated C atoms make the major contributions to the partially flat bands in the range of E$^{v}$ $\sim$ $-$3.5 to $-8$ eV.

The electronic properties of hydrogenated graphene are diversified by the chiral H-distribution. The chiral systems exhibit the narrow-gap or gapless behavior, as indicated in Figs. 4(d)-4(f). Two pairs of parabolic bands near E$_F$ mainly arise from the non-passivated C atoms; furthermore, the one, with the HOS and LUS at the M point, will determine a small direct gap. E$_g$'s are, respectively, 50 and 35 meV's for the 14 $\%$ and 3.6$\%$ H-concentrations (insets in Figs. 4(d) and 4(f)). As to another pair of parabolic bands, there are two critical points formed between K and $\Gamma$ points at $\pm$0.5 ($\pm$0.2) eV in the 14 $\%$ (3.6 $\%$) system. At the sufficiently low H-concentration, both chiral and armchair systems present the similar gapless band structures; that is, a pair of partially flat bands and a distorted Dirac-cone structure are revealed near E$_F$ simultaneously (Figs. 4(f) and 4(c)).

Up to now, ARPES can serve as a powerful experimental method in researching the main characteristics of energy bands of graphene systems. The Dirac-cone-like linear energy dispersions for graphene layers on SiC have been identified by ARPES \cite{ohta2008morphology}. Such measurements are utilized to verify the remarkable effects due to the number of layers \cite{ohta2007interlayer,sutter2009electronic}, stacking configurations \cite{ohta2006controlling}, gate voltages \cite{zhou2007substrate}, and adatom adsorptions \cite{zhou2008metal,schulte2013bandgap}. For example, a high-resolution direct observation shows that the alkali atoms on graphene surface can induce a lot of free electrons in the linear conduction band \cite{papagno2011large}.The hydrogenated graphenes are predicted to exhibit the distribution- and concentration-dependent electronic band structures. The feature-rich energy spectra, which include the opening of energy gap, the existence of the Dirac-cone structure, the anisotropic energy bands, the asymmetric valence and conduction bands, and the different energy dispersions near E$_F$, could be further test by using ARPES measurements.

How the spatial charge distributions induce the dramatic changes in energy bands can be fully comprehended by analyzing the charge density ($\rho$) and the difference of charge density ($\Delta$$\rho$) \cite{lin2015feature,lin2015h}. The former reveals the serious orbital hybridizations in C-C and C-H bonds. In pristine graphene (Fig. 5(a)), a strong covalent $\sigma$ bond, with high charge density of the (2$p_x$,2$p_y$,2s) orbitals, exists between two C atoms (rectangle in Fig. 5(a)). Such bonding is slightly reduced after hydrogenization (Figs. 5(b)-5(d)), so that the energy range of $\sigma$ bands is almost identical for any systems (Figs. 3 and 4). Also, 1s orbitals of H atoms strongly bond with 2$p_z$ orbitals of passivated C atoms (Figs. 5(b)-5(d)), being insensitive to H-concentration and thus leading to the similarly H-related bands (Figs. 3(b)-3(f) and 4). This indicates that 1s orbitals only partially hybridize with (2$p_x$,2$p_y$,2s) orbitals (discussed in Fig. 6).

The charge density difference is very useful in understanding the dependence of $\pi$ bonding on H-concentration. $\Delta$$\rho$ is created by substracting the charge density of isolated C and H atoms from that of a composite system. The weak $\pi$ bonding of graphene arising from the parallel 2$p_z$ orbitals is revealed at upper and lower surfaces (gray rectangle in Fig. 5(a)). All the $\pi$ bonds are terminated after the full hydrogenation (Fig. 5(b)), accounting for the absence of low-lying $\pi$ bands and the large band gap (Fig. 3(b)). The termination between the first- and second-nearest adjacent C atoms is clearly illustrated in $\Delta$$\rho$ (rectangle in Fig. 5(e)). With the decrease of H-concentration, the $\pi$ bonding is gradually recovered (Figs. 5(f)-5(h)). This is responds for the reduced band gap (Figs. 3(b)-3(f)). Similar spatial charge distributions are also observed by the chiral systems (Figs. 5(i) and 5(j)).

The main characteristics of DOS are determined by the band-edge states of energy dispersions; furthermore, the orbital-projected DOS directly reflects the feature-rich chemical bondings. The hydrogenated graphenes exhibit a lot of special structures (Van Hove singularities), as indicated in Figs. 6(a)-6(f). The delta-function-like peaks, the discontinuous shoulders and the logarithmically divergent peaks, respectively, arise from the partially flat bands, the extreme points and the saddle points of the parabolic bands. Furthermore, most of structures belong to the two latters. Concerning pristine graphene, DOS in the range of $-$3 $\le$ E $\le$ 3 eV is fully dominated by the $\pi$ bondings of the parallel 2$p_z$ orbitals (the red curve in Fig. 6(a)). It is vanishing at E$_{F}$=0 and linearly depends on energy, mainly owing to the isotropic Dirac-cone structure. And then, DOS presents a strong peak in the logarithmic form at $-2.2$ eV or $1.8$ eV, corresponding to the saddle M point (Fig. 3(a)). With the increasing energy ($\mid$E$\mid\,\ge\,3$ eV), the $\sigma$ bondings due to 2$p_x$ and 2$p_y$ orbitals start to make more important contributions to DOS, including an obvious shoulder structure at $-3$ eV and a strong symmetric peak at $-6.5$ eV. These two structures are caused by the local-maximum $\Gamma$ point and the saddle M point of the $\sigma$ bands, respectively. The quite different structures for the $\pi$ and $\sigma$ bondings clearly indicate the negligible interactions between 2$p_z$ and (2$p_x$, 2$p_y$) orbitals.

The energy and number of the special structures in DOS are thoroughly altered by the strong C-C and C-H bonds. For the fully hydrogenated graphene, DOS is vanishing within the range of $\sim\pm\,1.8$ eV, as shown in Fig. 6(b). A shoulder structure at $-1.8$ eV corresponds to the HOS due to the $\sigma$-electronic state of (2$p_x$, 2$p_y$) orbitals. The strong bondings of 2$p_{z}$ and 1$s$ orbitals lead to the similar peak structures in the orbital-projected DOS for $-3.5 \le$ E $\le -8$ eV (red and blue curves). With the decreasing H-concentration, the 25$\%$ zigzag system presents two prominent peaks at E $\sim \pm 1.3$ eV (Fig. 6(c)), in which they are associated with the weak $\pi$ bondings (Fig. 5(c)). The middle-energy DOS is also dominated by the 2$p_{x}$ and 2$p_{y}$ orbitals, while more peak structures arise from the increasing number of subbands. Specially, for the sufficiently low concentration, the main features of DOS are similar to those of pristine graphene in terms of the dominance of 2$p_{z}$ orbitals at low energy, a shoulder-like structure at $-3$ eV, and a strong symmetric peak at $-6.5$ eV, e.g., 2$\%$ zigzag system (Fig. 6(d)).

The strong chemical bondings cause the three distinct configurations to exhibit the similar structures in DOS (Figs. 6(b)-6(h), including the prominent peaks from the planar $\sigma$ bondings (Figs. 6(d), 6(f); 6(h)) and the bonding of 1$s$ and 2p$_{z}$ orbitals (Figs. 6(b), 6(e); 6(g)). On the other hand, the important differences among them lie in the low-energy DOS. For armchair system of 33$\%$ and chiral system of 14$\%$ (Figs. 6(e) and 6(g)), they possess two prominent peaks at E $\sim \pm 0.5$, which respectively come from two weakly dispersive bands and two critical points of parabolic bands (Figs. 4(a) and 4(d)). These two peaks are merged into a delta-function-like one at E$\sim 0$ at the low concentrations of 3.7$\%$ and 1.6$\%$ (Figs. 6(f) and 6(h)). Furthermore, their DOSs linearly depend on energy within E$\le \pm 1.1$ and E$\le \pm 0.7$ eV, respectively, and the main reason is the recovered distorted Dirac-cone structure. However, the significant difference between the gapless armchair system and narrow-gap chiral one is revealed by the DOS at E$\sim 0$ (Figs. 6(e) and 6(g)).

The STS measurement, in which the tunneling conductance (dI/dV) is approximately proportional to DOS and directly reflects its special structures, can provide an efficient way to confirm the distribution of H atoms. This method has been used for adatoms on graphene \cite{gyamfi2011fe}, carbon nanotubes \cite{wilder1998electronic}, graphene nanoribbons \cite{huang2012spatially}, and few-layer graphene \cite{choi2010atomic} . For example, it has been used to observe the Fermi-level shift and additional peaks near the Dirac point of graphene irradiated with Ar$^+$ ions \cite{tapaszto2008tuning}. Moreover, hydrogenated graphene that possesses a gap larger than $2$ eV has also been measured with this method. The main features in electronic properties, the concentration- and configuration-dependent energy gaps, the low-lying prominent peaks, and the H- and C-dominated peaks at middle energy, can be further investigated with STS. The STS measurements on the low- and middle-energy peaks can identify the configurations and concentrations of hydrogenated graphenes.\\

\vskip 0.6 truecm
\par\noindent
{\bf 4. Conclusion }
\vskip 0.3 truecm


The geometric and electronic properties of hydrogenated graphene are investigated by the DFT calculations. The chemical and physical properties are enriched by the configuration and concentration of adatoms. The strong H-C bonds and the drastic changes in C-C bonds dominate the optimized geometric structures, charge distributions, energy bands, and DOS. The C-C and H-C lengths, and the H-C-C angle are sensitive to the variation of H-concentration. The serious orbital hybridizations, strong $\sigma$ bonds, and weak $\pi$ bondings can diversify the electronic properties. There exist middle-gap, gapless, and narrow-gap systems, respectively, corresponding to zigzag, armchair, and chiral configurations. The main features of DOS are evidenced by the delta-function-like peaks, the discontinuous shoulders, and the logarithmically divergent peaks.

The hydrogen-enriched band structures exhibit the absence or recovery of low-lying $\pi$ bands, the weakly dispersive bands dominated by the second-nearest C atoms, the one pair of low-energy parabolic bands due to non-passivated C atoms, and the (C,H)-related partially flat bands. The band gap in 100 \% zigzag system is the largest one determined by the $\sigma$ bands, and it is largely reduced at lower concentrations. That the fully terminated $\pi$ bonding is replaced by the partially suppressed one is responsible for this result. As for armchair and chiral systems, the distorted Dirac cone structure is revealed at the sufficiently low concentration. The strong charge bondings of 2$p_z$ and 1s orbitals contribute to the partially flat bands at middle energy for all the systems, leading to the similar peaks of orbital-projected DOS. The decrease of H-concentration can induce the special structures in DOS, including the recovery of low-lying peaks, the emerged strong peaks, and the delta-funtion-like peak centered at $E_{F}$.

The hydrogenated graphene exhibit feature-rich properties, the optimal geometric structures, band gaps and energy dispersions, and special structures of DOS. They could be, respectively, examined by the STM, ARPES, and STS measurements, providing the useful informations about the hydrogen distribution and concentration. This system is suitable for the application of hydrogen storage. The tunable electronic properties are expected to be potentially important in nanoelectronic and nanophotonic devices.

\par\noindent {\bf Acknowledgments}

This work was supported by the Physics Division, National Center for Theoretical Sciences (South), the Nation Science Council of Taiwan (Grant No. NSC 102-2112-M-006-007-MY3). We also thank the National Center for High-performance Computing (NCHC) for computer facilities.

\newpage
\renewcommand{\baselinestretch}{0.2}

\newpage
\begin{table}[!hbp]
\begin{tabular}{ l l l l l l l l l l l }
\hline
\multirow{3}{*}{Type} & \multirow{3}{*}{$\#$H} & \multirow{3}{*}{$\#$C} & \multirow{3}{*}{H/C} & \multicolumn{4}{c}{ Bond length ({\AA})} & \multicolumn{2}{l}{Bond angle ($^{\circ}$)} & Shift on\\

 & & & & 1st- & 2nd(1)- & 2nd(2)- & \multirow{2}{*}{C-H} & H-C-C & C-C-C & x-axis({\AA}) \\
 & & & & \multicolumn{3}{c}{Nearest C-C} & & & & \\
\hline
\multirow{6}{*}{zigzag}
  & 0 & 2 & 0\% & 1.41 & 1.41 & 1.41 & - & 90 & 120 & 0 \\
  & 2 & 2 & 100\% & 1.53 & - & - & 1.12 & 107.1 & 111.8 & 0.22 \\
  & 2 & 8 & 25\% & 1.50 & 1.39 & - & 1.12 & 101.0 & 116.5 & 0.30 \\
  & 2 & 32 & 6.3\% & 1.49 & 1.39 & - & 1.13 & 104.2 & 114.2 & 0.57 \\
  & 2 & 50 & 4.0\% & 1.49 & 1.40 & - & 1.13 & 102.8 & 115.2 & 0.47 \\
  & 2 & 98 & 2.0\% & 1.49 & 1.40 & - & 1.13 & 103.1 & 115.0 & 0.56 \\
\hline
\multirow{3}{*}{armchair}
  & 2 & 6 & 33\% & 1.47 & 1.40 & - & 1.15 & 100.5 & 116.8 & 0.27 \\
  & 2 & 24 & 8.3\% & 1.48 & 1.41 & 1.40 & 1.14 & 101.6 & 116.1 & 0.39 \\
  & 2 & 54 & 3.7\% & 1.48 & 1.40 & 1.39 & 1.13 & 103.0 & 115.1 & 0.56 \\
\hline
\multirow{3}{*}{chiral}
  & 2 & 14 & 14\% & 1.50 & 1.43 & 1.36 & 1.12 & 104.6 & 113.9 & 0.46 \\
  & 2 & 56 & 3.6\% & 1.48 & 1.40 & 1.38 & 1.13 & 103.3 & 114.8 & 0.54 \\
  & 2 & 126 & 1.6\% & 1.48 & 1.40 & 1.40 & 1.13 & 101.5 & 116.1 & 0.49 \\
\hline
\end{tabular}
\caption{The calculated C-C and C-H bond lengths, H-C-C and C-C-C angles, and the shift on x-axis of passivated C for various concentrations.}
\end{table}

\newpage \centerline {\Large \textbf {FIGURE CAPTIONS}}

\vskip0.5 truecm 

 Fig. 1. Geometric structures of hydrogenated graphene with: (a) three classical periodical configurations, top view and side view of (b) fully hydrogenated graphene, (c) $25 \%$ zigzag configuration, (d) $33 \%$ armchair configuration, and (e) $14.3 \%$ chiral configuration.

 Fig. 2. The orbital hybridization of four orbitals (2$s$, 2$p_x$, 2$p_y$, 2$p_z$) and the hybridization of $sp^3$ and 1$s$ orbitals.

 Fig. 3. Band structures of hydrogenated graphenes: (a) pristine graphene, zigzag configurations with (b) $100 \%$, (c) $25 \%$, (d) $6.3 \%$, (e) $4 \%$, and (f) $2 \%$ H-concentrations.

 Fig. 4. Band structures of hydrogenated graphenes: armchair configurations with (a) $33 \%$, (b) $8.3 \%$, and (c) $3.7 \%$ H-concentrations, and chiral configurations with (d) $14.3 \%$, (e) $3.6 \%$, and (f) $1.6 \%$ H-concentrations. The insets are the low-lying band structures.

 Fig. 5. The charge density $\rho$ of (a) pristine graphene, (b) $100 \%$, (b) $25 \%$, and (c) $14.3 \%$ hydrogenated graphene. The charge density difference $\Delta\rho$ of hydrogenated graphene in (c) $100 \%$, (d) $25 \%$, (e) $6.3 \%$, (f) $4 \%$, (d) $14.3 \%$, and (d) $3.6 \%$ systems.

 Fig. 6. Orbital-projected DOS of (a) pristine graphene, zigzag configurations with (b) $100 \%$, (c) $25 \%$, and (d) $2 \%$ H-concentrations, and armchair configurations with (e) $33 \%$ and (f) $3.7 \%$ H-concentrations, and chiral configurations  with (g) $14.3 \%$, and (h) $3.6 \%$ H-concentrations.

\begin{figure}[htb]
\centering\includegraphics[width=16cm]{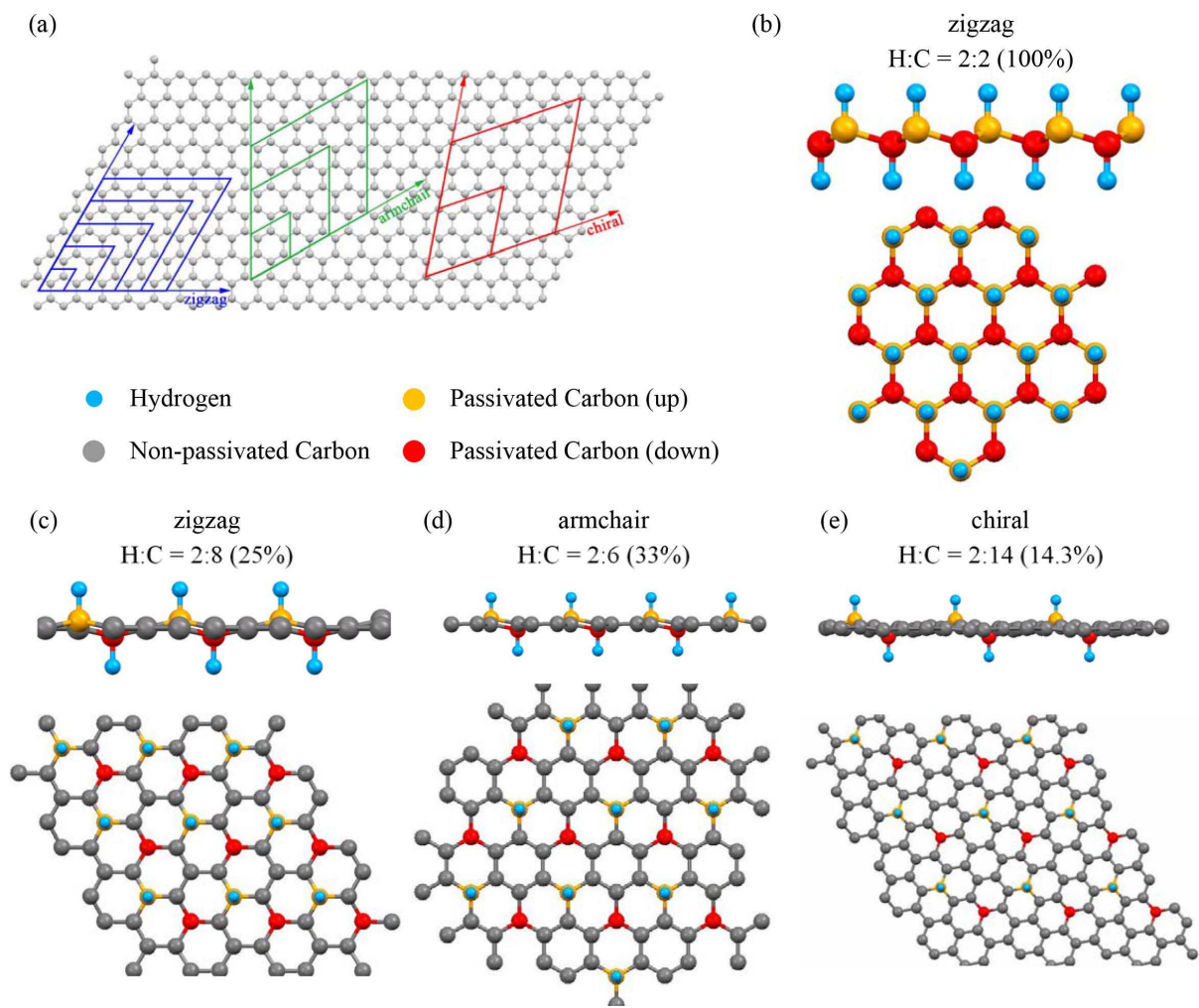}
\caption{Geometric structures of hydrogenated graphene with: (a) three classical periodical configurations, top view and side view of (b) fully hydrogenated graphene, (c) $25 \%$ zigzag configuration, (d) $33 \%$ armchair configuration, and (e) $14.3 \%$ chiral configuration.}
\end{figure}

\begin{figure}[htb]
\centering\includegraphics[width=12cm]{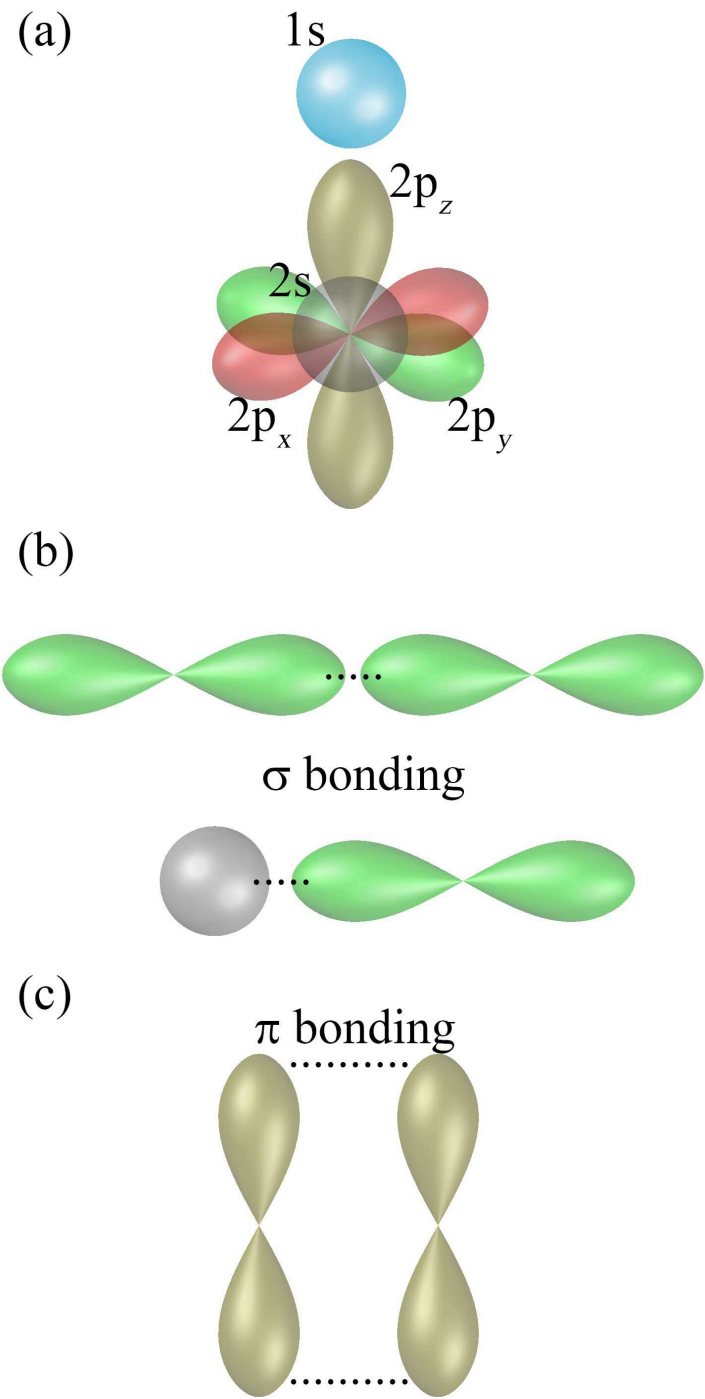}
\caption{The orbital hybridization of four orbitals (2$s$, 2$p_x$, 2$p_y$, 2$p_z$) and the hybridization of $sp^3$ and 1$s$ orbitals.}
\end{figure}

\begin{figure}[htb]
\centering\includegraphics[width=16cm]{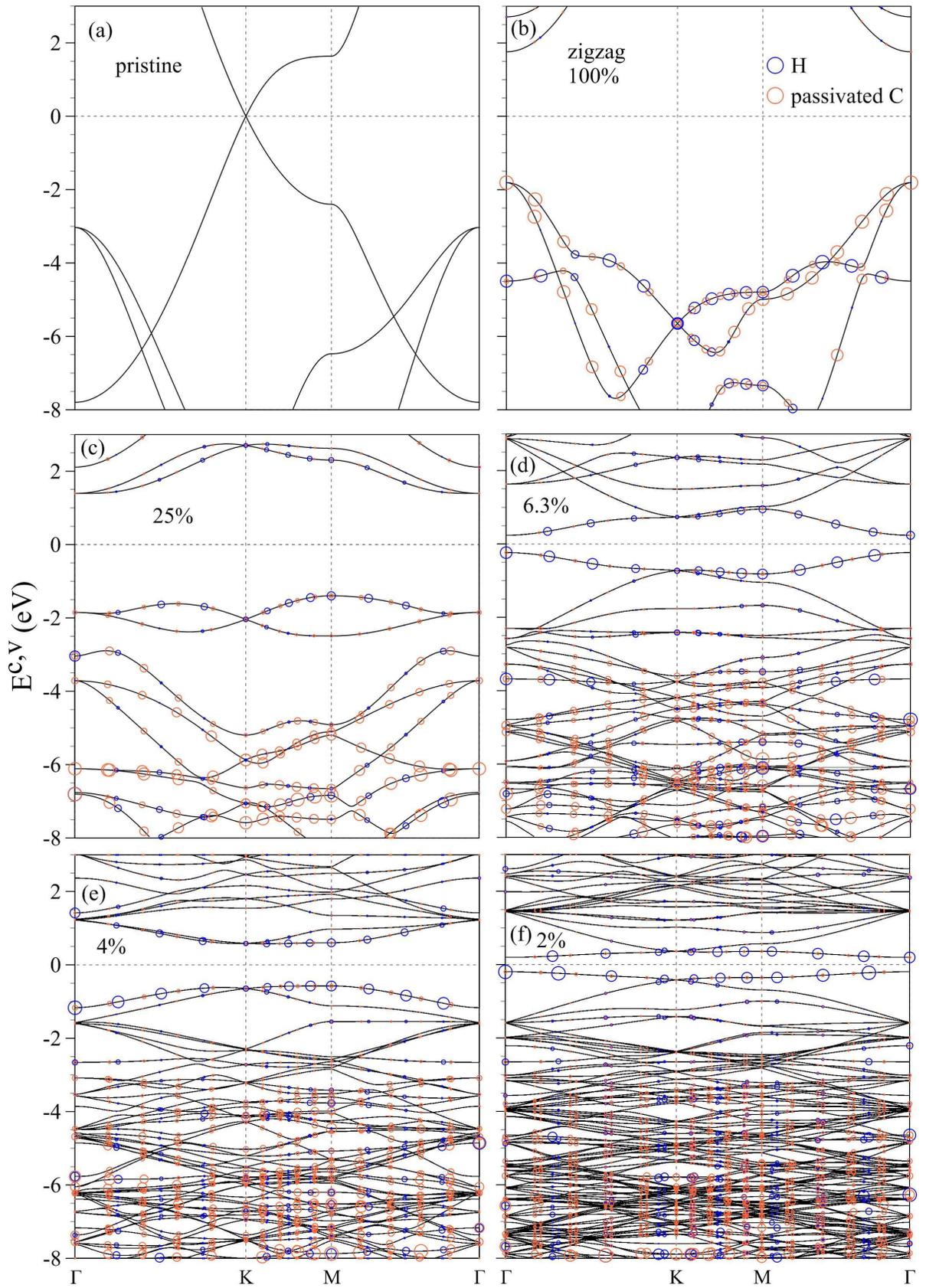}
\caption{Band structures of hydrogenated graphenes: (a) pristine graphene, zigzag configurations with (b) $100 \%$, (c) $25 \%$, (d) $6.3 \%$, (e) $4 \%$, and (f) $2 \%$ H-concentrations.}
\end{figure}

\begin{figure}[htb]
\centering\includegraphics[width=16cm]{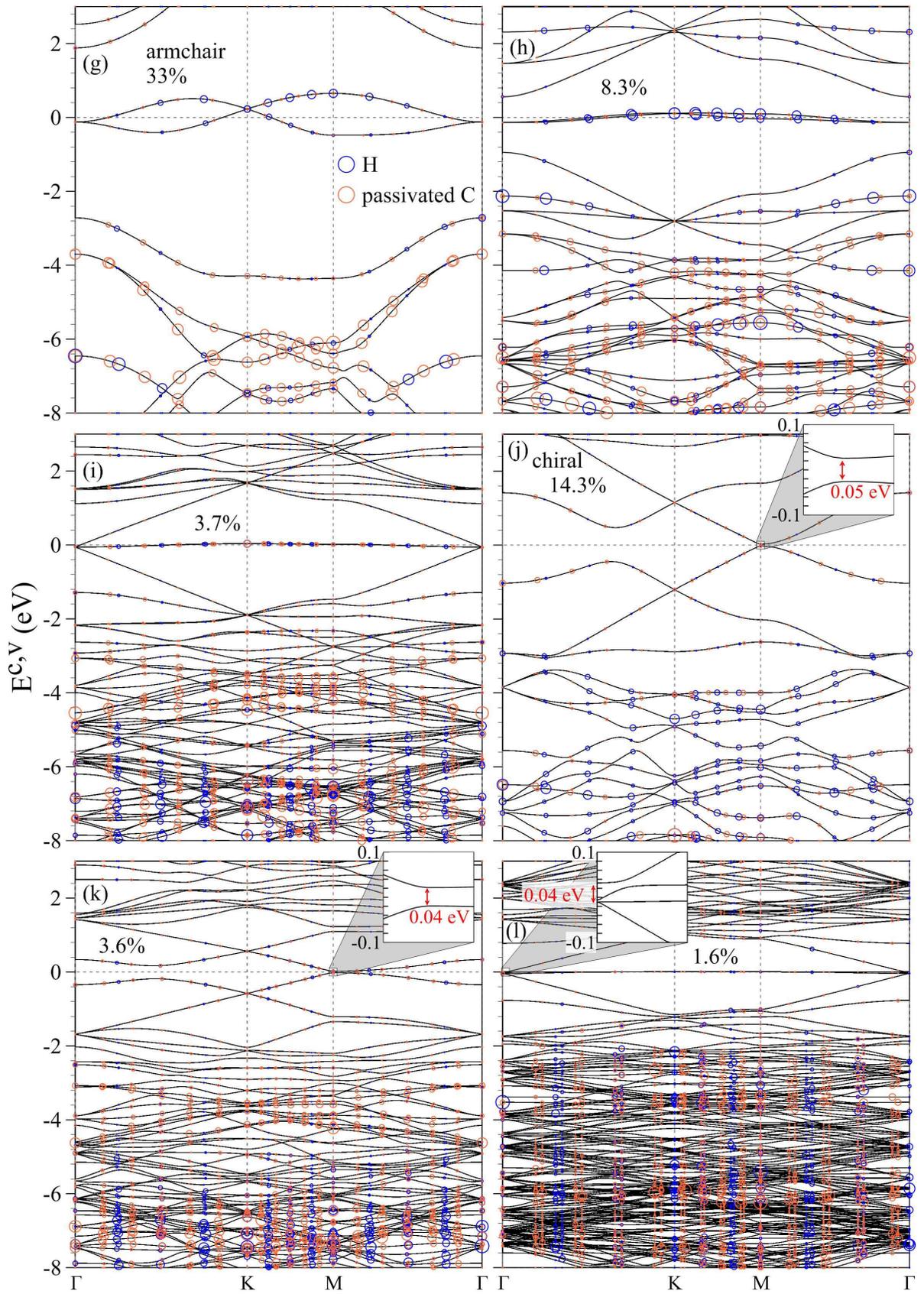}
\caption{Band structures of hydrogenated graphenes: armchair configurations with (a) $33 \%$, (b) $8.3 \%$, and (c) $3.7 \%$ H-concentrations, and chiral configurations with (d) $14.3 \%$, (e) $3.6 \%$, and (f) $1.6 \%$ H-concentrations. The insets are the low-lying band structures.}
\end{figure}

\begin{figure}[htb]
\centering\includegraphics[width=16cm]{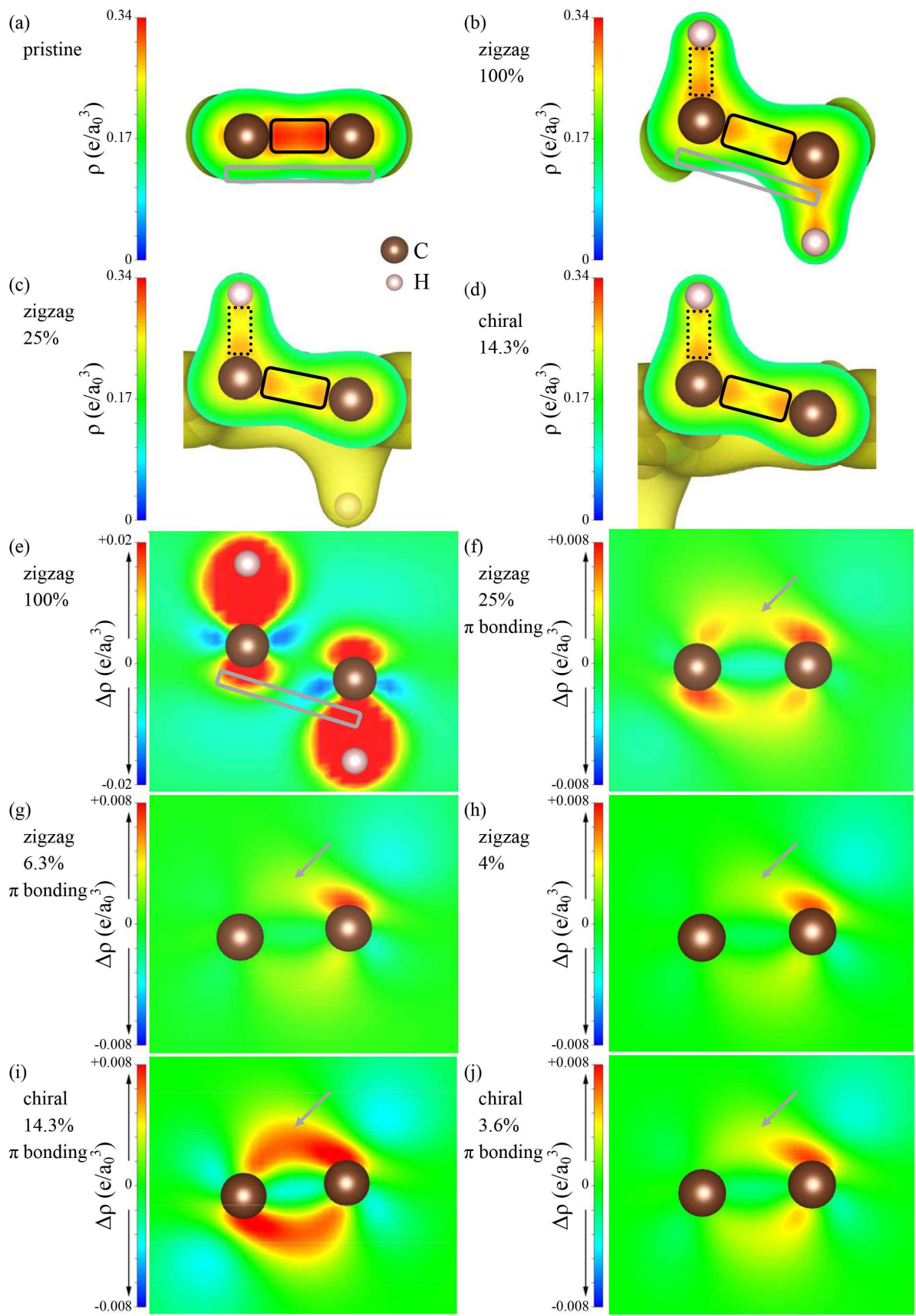}
\caption{The charge density $\rho$ of (a) pristine graphene, (b) $100 \%$, (b) $25 \%$, and (c) $14.3 \%$ hydrogenated graphene. The charge density difference $\Delta\rho$ of hydrogenated graphene in (c) $100 \%$, (d) $25 \%$, (e) $6.3 \%$, (f) $4 \%$, (d) $14.3 \%$, and (d) $3.6 \%$ systems.}
\end{figure}

\begin{figure}[htb]
\centering\includegraphics[width=16cm]{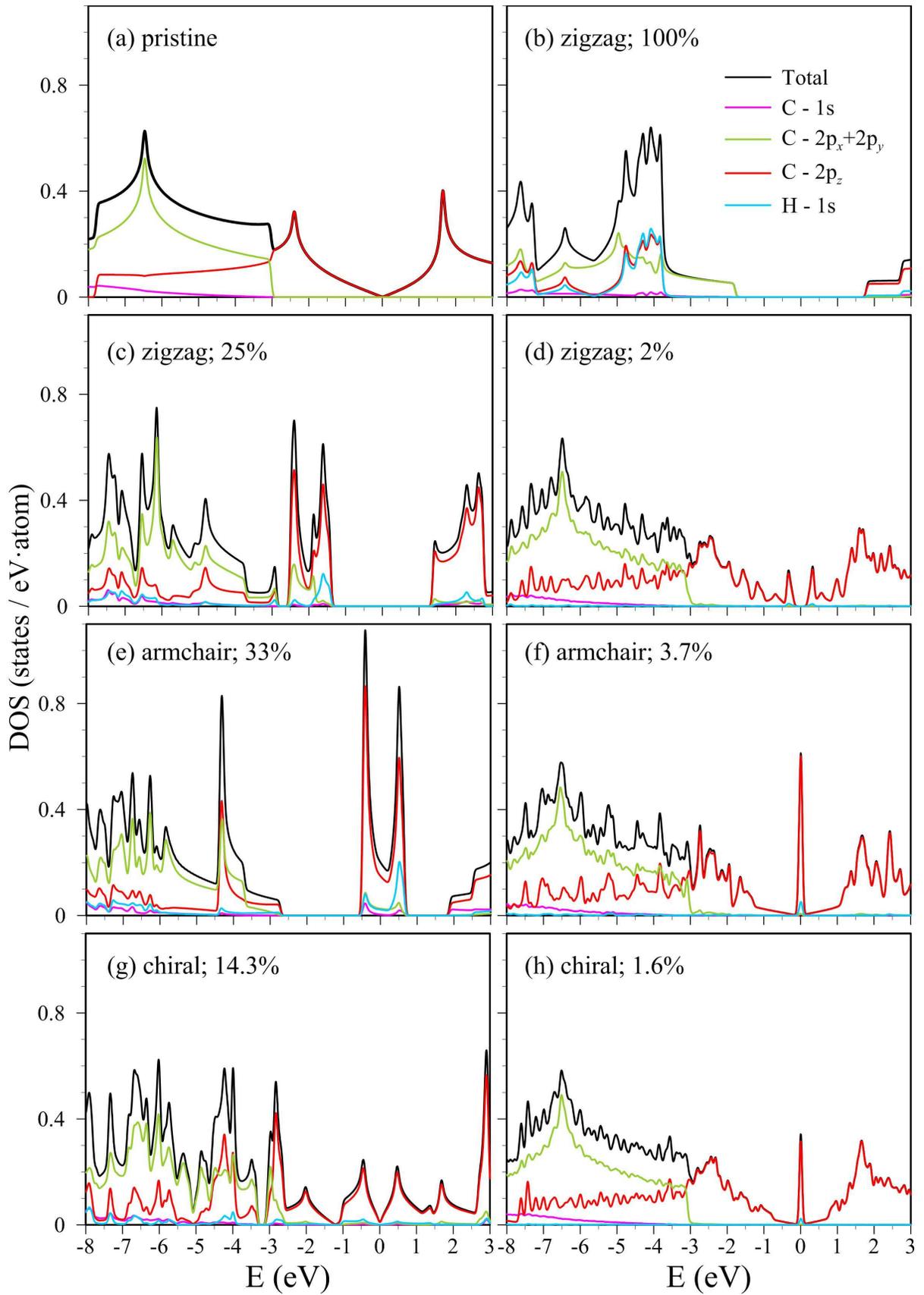}
\caption{Orbital-projected DOS of (a) pristine graphene, zigzag configurations with (b) $100 \%$, (c) $25 \%$, and (d) $2 \%$ H-concentrations, and armchair configurations with (e) $33 \%$ and (f) $3.7 \%$ H-concentrations, and chiral configurations  with (g) $14.3 \%$, and (h) $3.6 \%$ H-concentrations.}
\end{figure}

\end{document}